# Weak invariants in dissipative systems: Action principle and Noether charge for kinetic theory

Sumiyoshi Abe [1-4]

[1] *Department of Physics, College of Information Science and Engineering, Huaqiao University, Xiamen 361021, China*
[2] *Department of Physical Engineering, Mie University, Mie 514-8507, Japan*
[3] *Institute of Physics, Kazan Federal University, Kazan 420008, Russia*
[4] *ESIEA, 9 Rue Vesale, Paris 75005, France*

**Keywords**  kinetic theory, action principle, weak invariant, Noether charge

**Abstract**    In nonequilibrium classical thermostatistics, the state of a system may be described by not only dynamical/thermodynamical variables but also a kinetic distribution function. This "double structure" bears some analogy with that in quantum thermodynamics, where both dynamical variables and the Hilbert space are involved. Recently, the concept of weak invariants has repeatedly been discussed in the context of quantum thermodynamics. A weak invariant is defined in such a way that its value changes in time but its expectation value is conserved under time evolution prescribed by a kinetic equation. Here, a new aspect of a weak invariant is revealed for the classical Fokker-Planck equation as an example of classical kinetic equations. The auxiliary field formalism is applied to construction of the action for the kinetic equation. Then, it is shown that the auxiliary field is a weak invariant and is the Noether charge. The action is invariant under the transformation generated by the weak invariant. The result may shed light on possible roles of the symmetry principle in the kinetic descriptions of nonequilibrium systems.

## 1. Introduction

Quantum thermodynamics [1-3] is of importance for its possible contributions to classical thermodynamics in the nonequilibrium regime, in addition to its relevance to e.g. nanoscience and quantum information. A point is the fact that an element of a classical system is fundamentally described by a set of dynamical/thermodynamical variables, whereas in quantum mechanics the Hilbert space also accompanies. This "double structure" in quantum thermodynamics leads to diversity of the bath concepts. A typical example is the dephasing bath [4], the role of which is to realize decoherence of a quantum state and therefore has no classical counterparts. As long as classical equilibrium thermodynamics concerns, the bath simply implies the heat bath. In nonequilibrium classical thermodynamics, however, a situation becomes somewhat analogous to quantum thermodynamics if the kinetic approach is employed, where both dynamical/thermodynamical variables and time evolution of a probability distribution function of a system under consideration have to be treated.

Thermodynamic processes of interest are usually characterized by conserved values of quantities and variables. Examples are the isothermal, isochoric, isobaric and isoentropic processes. In a couple of recent works [5,6] on finite-time quantum thermodynamics, a process termed the isoenergetic process has been studied by the use of the Lindblad equation [7,8]. Along such a process, the internal energy is kept constant. This requires the concept of the energy bath with no counterparts in equilibrium classical thermodynamics and may be realized by the energy transfer technique [9,10]. Such an exotic process manifests the implications of the quantum-mechanical violation of the law of equipartition of energy to thermodynamics since it is different from the isothermal process. There, although the Hamiltonian of an objective subsystem explicitly depends on time and therefore its spectrum is not constant in time, the internal energy, i.e. its expectation value, remains constant. This condition turns out to be able to determine the Lindbladian operators without recourse to detailed knowledge about an interaction between the subsystem and the energy bath [5,6].

In general, a time-dependent observable is referred to as a weak invariant [11] if its spectrum depends on time but the expectation value is conserved in the course of time evolution described by a master equation such as the Lindblad equation. The time-dependent Hamiltonians in the isoenergetic processes discussed in [5,6] are thus weak invariants. In contrast, a time-dependent observable with the time-independent spectrum is to be called a strong invariant. The Lewis-Riesenfeld invariant [12] associated with a time-dependent Hamiltonian is an example of strong invariants. It has been applied to a wide variety of problems including quantum optics and related systems [13-15], nonstationary quantum field theory [16-18], geometric phases [19-21] quantum computation [22] and quantum cosmology [23]. It is worth emphasizing that the classical counterpart of the Lewis-Riesenfeld strong invariant can be derived by Noether's theorem [24] (see also [25] for a simplified discussion). Its generalization to dissipative systems has recently been presented in [11], where the weak invariant of the time-dependent quantum damped oscillator has explicitly been constructed by the use of the Lindblad equation.

The action principle for kinetic theory has repeatedly been discussed in the literature. Among others, the approach [26, 27] should be noticed, in particular. There, an auxiliary field is introduced in order for construction of the action functional for the Liouville-von Neumann equation, and thus this approach is referred to as the auxiliary field formalism. A primary reason for the introduction of such an extra field is due to the fact that the Liouville-von Neumann equation is a first-order differential equation in time for a density matrix. Although the Liouville-von Neumann equation possesses time-reversal invariance (for autonomous isolated systems), more general master equations may violate such symmetry. For example, the Lindblad equation repeatedly mentioned above describes dissipative dynamics. Then, the auxiliary field formalism comes to play a prominent role: the action can be constructed for dissipative kinetics through the extension of the space of variables. It is also mentioned that this formalism offers a powerful too for obtaining approximate solutions in analytic forms if combined with the Rayleigh-Ritz method [27,28].

Recently, the auxiliary field formalism has been applied to classical fractional kinetic equations [28,29] as well as a general quantum master equation [30]. It has been found that the auxiliary field introduced in the action may actually be a weak invariant of the corresponding master equation. This fact indicates the existence of a certain deeper structure that must be related to the symmetry principle behind kinetic theory. Understanding such a principle may, in turn, cast fresh light on nonequilibrium thermodynamics even at the classical level.

In this paper, the action principle and weak invariants are studied in the context of classical kinetic theory. Although a specific form of a kinetic equation is not needed as long as it is linear, the Fokker-Planck equation [31,32] is employed as an important example for the sake of clarity. For the Fokker-Planck equation is derived from the Langevin equation that describes the irreversible relaxation processes [33], and therefore it is of direct relevance to nonequilibrium thermodynamics. In addition, the weak invariant of the Fokker-Planck equation has already been discussed in [34]. Now, the primary purpose of the present work is to show how the auxiliary field as the weak invariant in the action principle plays a role of the Noether charge and accordingly the action $I$ satisfies

$$\Delta_\varepsilon I = 0, \qquad (1.1)$$

where $\Delta_\varepsilon$ stands for the infinitesimal transformation generated by the weak invariant. This result may give new insight into possible roles of the symmetry principle in kinetic theory of dissipative systems and associated nonequilibrium thermodynamic processes.

The present paper is organized as follows. In Section 2, the concept of weak invariants is explained for the Fokker-Planck equation, and the action principle for the equation is formulated by the use of the auxiliary field. Then, the auxiliary field is shown to be a weak invariant. In Section 3, the canonical formulation is developed, and the action is shown to be invariant under the transformation generated by the auxiliary field as a weak invariant. Thus, the weak invariant is found to be the Noether charge. Section 4 is devoted to concluding remarks.

## 2. Weak invariant and action principle for Fokker-Planck equation

Consider the Fokker-Planck equation of the following form [31,32]:

$$\frac{\partial P(x,t)}{\partial t} = -\frac{\partial}{\partial x}\left[K(x,t)P(x,t)\right] + \frac{\partial^2}{\partial x^2}\left[D(x,t)P(x,t)\right]$$

$$\equiv L_t\left[P(x,t)\right]. \tag{2.1}$$

Here, $P(x,t)dx$ is the probability of finding the value of a physical variable $x$ in the interval $[x, x+dx]$ at time $t$. The domain of the probability distribution function is taken to be $(-\infty, \infty) \times [t_i, t_f]$. The operator $L_t$ is allowed to possess time dependence through $K$ in the drift term and the positive diffusion coefficient $D$. Generalization to the case of many variables is straightforward. Equation (2.1) does not possess time-reversal invariance, in general. The quantity conserved under time evolution is nothing but the total probability $\int_{-\infty}^{\infty} dx\, P(x,t)$ that may be set equal to unity.

A quantity, $J(x,t)$, is said to be a weak invariant associated with equation (2.1) if it satisfies

$$\frac{\partial J(x,t)}{\partial t} + L_t^\dagger\left[J(x,t)\right] = 0, \tag{2.2}$$

where $L_t^\dagger \equiv K(x,t)\,\partial/\partial x + D(x,t)\,\partial^2/\partial x^2$ is the adjoint of $L_t$ in equation (2.1). Then, it can immediately be ascertained that the expectation value

$< J > = \int_{-\infty}^{\infty} d x \, J(x,t) \, P(x,t)$ is conserved:

$$\frac{d < J >}{d t} = 0. \qquad (2.3)$$

Here, the probability distribution function and its spatial derivative need be required to decrease at the spatial infinities sufficiently rapidly. On the other hand, a weak invariant can be free from boundary conditions as long as its expectation value is finite and equation (2.3) holds. Accordingly, there exist large degrees of freedom in its choice. This is certainly an advantageous point since it implies that a wide class of quantities satisfying equation (2.2) can be chosen as a weak invariant, depending on physical interest. However, as seen in the subsequent discussion, a weak invariant is subject to the final condition, not the initial condition.

Although the first moment of the weak invariant satisfies equation (2.3), the higher moments do not, in general. In fact, the variance $(\Delta J)^2 = < J^2 > - < J >^2$ monotonically increases as follows:

$$\frac{d (\Delta J)^2}{d t} = 2 < D (J')^2 >, \qquad (2.4)$$

where $J' \equiv \partial J(x,t) / \partial x$.

Because of its irreversible nature and the structure of the first-order time derivative, it is nontrivial to develop the action principle for the Fokker-Planck equation. A quantum-mechanical analog of this issue has been studied in [26,27]. There, an auxiliary field $\Lambda$ has been introduced for the extension of the space of variables, by which the action can be constructed. In the present case, the action for equation (2.1) reads

$$I[P, \Lambda] = \int_{t_i}^{t_f} dt \int_{-\infty}^{\infty} dx \, \pounds - \frac{1}{2} \int_{-\infty}^{\infty} dx \left( \Lambda P \big|_{t=t_i} + \Lambda P \big|_{t=t_f} \right), \qquad (2.5)$$

where $\pounds$ is the Lagrangian density given by

$$\pounds = \frac{1}{2}\left( \Lambda \frac{\partial P}{\partial t} - \frac{\partial \Lambda}{\partial t} P \right) - \Lambda \left( -\frac{\partial}{\partial x}(KP) + \frac{\partial^2}{\partial x^2}(DP) \right). \qquad (2.6)$$

The second term on the right-hand side in equation (2.4) is the temporal boundary one that is peculiar in the auxiliary field formalism [26,27]. Upon taking the variation of the action, one should take into account that $P(x,t)$ is a probability distribution function satisfying the normalization condition, which needs be treated as a constraint. However, it is actually not necessary to add such a constraint to the action, since it can be eliminated by the redefinition of the auxiliary field $\Lambda(x,t)$. Indeed, if the auxiliary field is redefined as

$$\Lambda(x,t) \rightarrow \Lambda(x,t) + \int_{t}^{t_f} ds\, \lambda(s), \qquad (2.7)$$

then, the term

$$\int_{t_i}^{t_f} dt\, \lambda(t) \left( \int_{-\infty}^{\infty} dx\, P(x,t) - \int_{-\infty}^{\infty} dx\, P(x,t_i) \right) \qquad (2.8)$$

automatically appears in the action. This is nothing but the constraint on the normalization condition, if the initial probability distribution function is normalized. And, $\lambda(t)$ plays a role of the Lagrange multiplier. It is noted that equation (2.7) does not change the final condition $\Lambda(x, t_f)$.

The variations of the action in equation (2.5) with respect to $\Lambda(x,t)$ and $P(x,t)$ yield

$$\delta_\Lambda I[P, \Lambda] = \int_{t_i}^{t_f} dt \int_{-\infty}^{\infty} dx \left( \frac{\partial P}{\partial t} + \frac{\partial}{\partial x}(KP) - \frac{\partial^2}{\partial x^2}(DP) \right) \delta\Lambda - \int_{-\infty}^{\infty} dx\, P\, \delta\Lambda \bigg|_{t=t_f}, \quad (2.9)$$

$$\delta_P I[P, \Lambda] = -\int_{t_i}^{t_f} dt \int_{-\infty}^{\infty} dx \left( \frac{\partial \Lambda}{\partial t} + K\frac{\partial \Lambda}{\partial x} + D\frac{\partial^2 \Lambda}{\partial x^2} \right) \delta P - \int_{-\infty}^{\infty} dx\, \Lambda\, \delta P \bigg|_{t=t_i}, \quad (2.10)$$

respectively.

From equation (2.9), the Fokker-Planck equation (2.1) is derived if the final condition $\delta\Lambda(x, t_f) = 0$ is imposed, i.e. the final condition on the auxiliary field being fixed. In this respect, recall also that equation (2.7) keeps $\Lambda(x, t_f)$ unchanged.

On the other hand, the equation for the auxiliary field

$$\frac{\partial \Lambda(x,t)}{\partial t} + L_t^\dagger [\Lambda(x,t)] = 0 \quad (2.11)$$

is obtained from equation (2.10) under the initial condition $\delta P(x, t_i) = 0$, i.e. $P(x, t_i)$

being fixed. This equation shows that the spatial integral of the auxiliary field is not necessarily a conserved quantity, in contrast to the probability distribution function.

A crucial point is the fact that equations (2.2) and (2.11) are identical to each other. It is therefore concluded that *the auxiliary field introduced into the action for the Fokker-Planck equation is a weak invariant associated with the equation*. In this way, the auxiliary field, which used to be a mathematical object introduced rather formally, is now endowed with a physical significance.

## 3. Weak Invariant as Noether charge

Canonical theory is constructed from the Lagrangian density in equation (2.6). The canonical momentum densities conjugate to the probability distribution function and auxiliary field are: $\Pi_P = \partial \mathcal{L} / \partial (\partial P / \partial t) = \Lambda / 2$ and $\Pi_\Lambda = \partial \mathcal{L} / \partial (\partial \Lambda / \partial t) = -P / 2$, respectively. $\partial P / \partial t$ and $\partial \Lambda / \partial t$ cannot be solved in terms of the canonical momentum densities. Accordingly, the following constraints are present: $\chi_1 \equiv \Pi_P - \Lambda / 2 \approx 0$, $\chi_2 \equiv \Pi_\Lambda + P / 2 \approx 0$, where the symbol "$\approx$" denotes the weak equality in Dirac's notation [35]. The equal-time Poisson bracket is defined in terms of the functional derivatives as follows: $\{A(t), B(t)\} = \int_{-\infty}^{\infty} dx \sum_{i=1,2} \left( \delta A(t) / \delta Q_i(x,t) \cdot \delta B(t) / \delta \Pi_i(x,t) - \delta A(t) / \delta \Pi_i(x,t) \cdot \delta B(t) / \delta Q_i(x,t) \right)$. In this expression, $Q_1(x,t)$, $Q_2(x,t)$, $\Pi_1(x,t)$ and $\Pi_2(x,t)$ denote $P(x,t)$, $\Lambda(x,t)$, $\Pi_P(x,t)$ and $\Pi_\Lambda(x,t)$, respectively. In general, $A(t)$ and $B(t)$ are functionals that are the spatial integrals of functions of these four fields. Then, the basic equal-time Poisson-bracket relations are: $\{P(x,t), \Pi_P(x',t)\} = \delta(x-x')$, $\{\Lambda(x,t), \Pi_\Lambda(x',t)\} = \delta(x-x')$, and all the other combinations vanish. Therefore, it follows that $\{\chi_i(x,t), \chi_j(x',t)\} = -\varepsilon_{ij}\delta(x-x')$, where $\varepsilon_{ij}$ is the Levi-Civita symbol in two dimensions (i.e., $\varepsilon_{12} = -\varepsilon_{21} = 1$, $\varepsilon_{11} = \varepsilon_{22} = 0$). So, the constraints are of the second-class [35]. A standard way of eliminating them is to use the Dirac bracket: $\{A(t), B(t)\}_D = \{A(t), B(t)\}$ $-\sum_{i,j=1,2}\int_{-\infty}^{\infty}\int_{-\infty}^{\infty} dy\,dy'\{A(t), \chi_i(y,t)\}C_{ij}(y,y')\{\chi_j(y',t), B(t)\}$, where $C_{ij}(y,y')$ is the quantity satisfying $\sum_{k=1,2}\int_{-\infty}^{\infty} dy\,\{\chi_i(x,t), \chi_k(y,t)\}C_{kj}(y,y') = \delta_{ij}\delta(x-y')$ and is found to be $C_{ij}(y,y') = \varepsilon_{ij}\delta(y-y')$. The second-class constraints become the identical equalities in this formalism. Then, the basic equal-time Dirac-bracket relations are given as follows:

$$\{P(x,t), \Lambda(x',t)\}_D = \delta(x-x'), \qquad \{P(x,t), P(x',t)\}_D = 0,$$

$$\{\Lambda(x,t), \Lambda(x',t)\}_D = 0. \tag{3.1}$$

The first relation implies that the probability distribution function and auxiliary field are canonically conjugate to each other. It should be noted that the first relation never contradicts the normalization condition on the probability distribution function since such a condition is, as mentioned after equation (2.6), a constraint (i.e. a weak equation): $\int_{-\infty}^{\infty} dx\,P(x,t) \approx \int_{-\infty}^{\infty} dx\,P(x,t_i)$.

The Hamiltonian density is given by the Legendre transformation of the Lagrangian density in equation (2.6). Its spatial integration is the Hamiltonian and is explicitly given by $H(t) = \int_{-\infty}^{\infty} dx\, \Lambda(x,t)\, L_t[P(x,t)] = \int_{-\infty}^{\infty} dx\, L_t^{\dagger}[\Lambda(x,t)]\, P(x,t)$. Then, equations (2.1) and (2.11) are rewritten as $\partial P(x,t)/\partial t = \{P(x,t), H(t)\}_D$ and $\partial \Lambda(x,t)/\partial t = \{\Lambda(x,t), H(t)\}_D$, respectively. Thus, the Fokker-Planck equation is recast in the Liouvillian form.

The primary purpose of the present work is to show that the action is invariant under the transformation generated by the weak invariant. For this, it is convenient to set up the non-equal-time Dirac-bracket relations. Such relations are obtained as follows. Solve equations (2.1) and (2.11) formally as

$$P(x,t) = T \exp\left(\int_{t'}^{t} ds\, L_s\right) P(x,t'), \qquad (3.2)$$

$$\Lambda(x,t') = T \exp\left(-\int_{t}^{t'} ds\, L_s^{\dagger}\right) \Lambda(x,t), \qquad (3.3)$$

respectively, where $T$ stands for the chronological symbol responsible for ordering of the product of operators noncommutative at different times [36]. From these, the relations in equation (3.1) are generalized as follows:

$$\{P(x,t), \Lambda(x',t')\}_D = T \exp\left(-\int_{t}^{t'} ds\, L'^{\dagger}_s\right) \delta(x-x'), \qquad \{P(x,t), P(x',t')\}_D = 0,$$

$$\{\Lambda(x,t), \Lambda(x',t')\}_D = 0, \qquad (3.4)$$

where $L'^{\dagger}_t$ denotes $L^{\dagger}_t$ in equation (2.2) with respect to the variable $x'$.

Now, the invariant nature of the action can be made manifest. The transformation of a certain physical quantity, say $Q$, generated by the weak invariant $\Lambda$ is defined by

$$\Delta_\varepsilon Q = \{Q, G_\varepsilon\}_D, \tag{3.5}$$

where

$$G_\varepsilon \equiv \int_{-\infty}^{\infty} dx\, \varepsilon(x,t)\, \Lambda(x,t) \tag{3.6}$$

with $\varepsilon(x,t)$ being an arbitrary infinitesimal function on a compact support so that the integral in equation (3.6) converges. Then, taking the action in equation (2.5) as $Q$, one finds

$$\Delta_\varepsilon I = \{I, G_\varepsilon\}_D$$

$$= -\int_{t_i}^{t_f} dt \int_{-\infty}^{\infty} dx \left( \frac{\partial \Lambda(x,t)}{\partial t} + L^{\dagger}_t[\Lambda(x,t)] \right)$$

$$\times \int_{-\infty}^{\infty} dx'\, \varepsilon(x',t')\, T\exp\left(-\int_t^{t'} ds\, L'^{\dagger}_s\right) \delta(x-x')$$

$$- \int_{-\infty}^{\infty} dx\, \Lambda(x,t_i)\, \Delta_\varepsilon P(x,t_i). \tag{3.7}$$

Since the initial probability distribution function is fixed, the last term vanishes.

Therefore, from equation (2.11), it follows that equation (1.1) in fact holds. Consequently, the action is seen to be invariant under the transformation generated by the weak invariant. This result offers a reinterpretation of the variation in equation (2.10) in terms of the symmetry principle.

## 4. Concluding remarks

The action principle has been studied for the Fokker-Planck equation, and the associated weak invariant has been shown to be the Noether charge generating the transformation, under which the action remains unchanged.

It is noted that the variational approach discussed in the present work is radically different from the one that is directly applied to thermodynamic variables and is somewhat criticized in [37].

The theory developed here is free from a specific form of a kinetic equation (at least, as long as it is linear). For example, it is straightforward to apply it to fractional kinetics with spatiotemporal nonlocality. In fact, the action principle for the fractional Fokker-Planck equation and the corresponding canonical theory have already been constructed [29].

It is also pointed out that there have been continuous research activities of nonlocal nonequilibrium thermodynamics [38-41]. It may be of interest to examine such discussions from the viewpoint of generalized kinetics as in the traditional shift from the kinetic stage to the hydrodynamic one. Weak invariants and their possible roles in generalized kinetics are yet to be explored.

As long as limited to genuine classical thermodynamics, the bath implies the heat bath even in the nonequilibrium regime, and accordingly exotic baths such as the energy bath mentioned in Section 1 do not exist, there. However, the thermodynamic formalism has high-level universality. Thermodynamic analogs are ubiquitously observed in various fields such as econophysics, physics of seismicity, or time series analysis, in general. Nontraditional approaches including the present are expected to widen the realm of the thermodynamic formalism.

**Acknowledgment.** The author would like to thank Congjie Ou and Péter Ván for discussions and their interest in the present work.

**Funding Statement.** This work has been supported by a grant from National Natural Science Foundation of China (No. 11775084), the program of Fujian Province, China, and the program of Competitive Growth of Kazan Federal University from the Ministry of Education and Science, Russian Federation.

**Data Accessibility.** This article has no additional data.

**Competing interest.** The author declares no competing interests.

**References**

1. Gemmer J, Michel M, Mahler G. 2009 *Quantum thermodynamics*. 2nd. edn. Berlin, Germany: Springer-Verlag.
2. Mahler G. 2015 *Quantum thermodynamic processes*. Singapore: Pan Stanford.
3. Kosloff R. 2013 Quantum thermodynamics: a dynamical viewpoint. *Entropy* **15**, 2100-2128. (doi:10.3390/e15062100)
4. Giusteri GG, Recrosi F, Schaller G, Celardo GL. 2017 Interplay of different environments in open quantum systems: breakdown of the additive approximation.

*Phys. Rev. E* **96**, 012113. (doi:10.1103/PhysRevE.96.012113)

5. Ou C, Abe S. 2019 Weak invariants, temporally local equilibria and isoenergetic processes described by the Lindblad equation. *EPL* **125**, 60004. (doi:10.1209/0295-5075/125/60004)

6. Ou C, Yokoi Y, Abe S. 2019 Spin isoenergetic process and the Lindblad equation. *Entropy* **21**, 503. (doi:10.3390/e21050503)

7. Lindblad G. 1976 On the generators of quantum dynamical semigroups. *Commun. Math. Phys.* **48**, 119-130. (doi:10.1007/BF01608499)

8. Gorini V, Kossakowski A, Sudarshan ECG. 1976 Completely positive dynamical semigroups of *N*-level systems. *J. Math. Phys.* **17**, 821-825. (doi:10.1063/1.522979)

9. Gorman DJ, Hemmerling B, Megidish E, Moeller SA, Schindler P, Sarovar M, Haeffner H. 2018 Engineering vibrationally assisted energy transfer in a trapped-ion quantum simulator. *Phys. Rev. X* **8**, 011038. (doi:10.1103/PhysRevX.8.011038)

10. Henao I, Serra RM. 2018 Role of quantum coherence in the thermodynamics of energy transfer. *Phys. Rev. E* **97**, 062105. (doi:10.1103/PhysRevE.97.062105)

11. Abe S. 2016 Weak invariants of time-dependent quantum dissipative systems. *Phys. Rev. A* **94**, 032116. (doi:10.1103/PhysRevA.94.032116)

12. Lewis Jr. HR, Riesenfeld WB. 1969 An exact quantum theory of the time-dependent harmonic oscillator and of a charged particle in a time-dependent electromagnetic field. *J. Math. Phys.* **10**, 1458-1473. (doi:10.1063/1.1664991)

13. Hartley JG, Ray JR. 1982 Coherent states for the time-dependent harmonic oscillator. *Phys. Rev. D* **25**, 382-386. (doi:10.1103/PhysRevD.25.382)

14. Rajagopal AK, Marshall JT. 1982 New coherent states with applications to time-dependent systems. *Phys. Rev. A* **26**, 2977-2980. (doi:10.1103/PhysRevA.26.2977)

15. Abe S, Ehrhardt R. 1993 Effects of anharmonicity on nonclassical states of the time-dependent harmonic oscillator. *Phys. Rev. A* **48**, 986-994. (doi:10.1103/PhysRevA.48.986)

16. Abe S, Ehrhardt R. 1993 Method of invariant in nonstationary field theory. *Z. Phys. C* **57**, 471-474. (doi:10.1007/BF01474341)

17. Bertoni C, Finelli F, Venturi G. 1998 Adiabatic invariants and scalar fields in

a de Sitter space-time. *Phys. Lett. A* **237**, 331-336.
(doi:10.1016/S0375-9601(97)00707-X)

18. Gao XC, Fu J, Shen JQ. 2000 Quantum-invariant theory and the evolution of a Dirac field in Friedmann–Robertson–Walker flat space-times. *Eur. Phys. J. C* **13**, 527-541. (doi:10.1007/s100520000257)

19. Datta N, Ghosh G, Engineer MH. 1989 Exact integrability of the two-level system: Berry's phase and nonadiabatic corrections. *Phys. Rev. A* **40**, 526-529. (doi:10.1103/PhysRevA.40.526)

20. Dittrich W, Reuter M. 1991 Berry phase contribution to the vacuum persistence amplitude; effective action approach. *Phys. Lett. A* **155**, 94-98. (doi:10.1016/0375-9601(91)90571-O)

21. Gao XC, Xu JB, Qian TZ. 1991 Geometric phase and the generalized invariant formulation. *Phys. Rev. A* **44**, 7016-7021. (doi:10.1103/PhysRevA.44.7016)

22. Sarandy MS, Duzzioni EI, Serra RM. 2011 Quantum computation in continuous time using dynamic invariants. *Phys. Lett. A* **375**, 3343-3347. (doi:10.1016/j.physleta.2011.07.041)

23. Abe S. 1993 Fluctuations around the Wheeler-DeWitt trajectories in third-quantized cosmology. *Phys. Rev. D* **47**, 718-721. (doi:10.1103/PhysRevD.47.718)

24. Lutzky M. 1978 Noether's theorem and the time-dependent harmonic oscillator *Phys. Lett. A* **68**, 3-4. (doi:10.1016/0375-9601(78)90738-7)

25. Abe S, Itto Y, Matsunaga M. 2009 On Noether's theorem for the invariant of the time-dependent harmonic oscillator. *Eur. J. Phys.* **30**, 1337-1343. (doi:10.1088/0143-0807/30/6/011)

26. Balian R, Vénéroni M. 1981 Time-dependent variational principle for predicting the expectation value of an observable. *Phys. Rev. Lett.* **47**, 1353-1356. (doi:10.1103/PhysRevLett.47.1353); Erratum. *ibid.* **47**, 1765. (doi:10.1103/PhysRevLett.47.1765.3)

27. Éboli O, Jackiw R, Pi S.-Y. 1988 Quantum fields out of thermal equilibrium. *Phys. Rev. D* **37**, 3557-3581. (doi:10.1103/PhysRevD.37.3557)

28. Abe S. 2013 Variational principle for fractional kinetics and the Lévy Ansatz. *Phys. Rev. E* **88**, 022142. (doi:10.1103/PhysRevE.88.022142)

29. Abe S. 2018 Hamiltonian formulation of fractional kinetics. *Eur. Phys. J. Special*

*Topics* **227**, 683-691. (doi:10.1140/epjst/e2018-00123-8)

30. Abe S, Ou C. 2019 Action principle and weak invariants. *Results in Physics* **14**, 102333. (doi:10.1016/j.rinp.2019.102333)
31. Risken H. 1989 *The Fokker-Planck equation*. 2nd. edn. Berlin, Germany: Springer-Verlag.
32. Cáceres MO. 2017 *Non-equilibrium statistical physics with application to disordered systems*. Cham, Switzerland: Springer International.
33. Onsager L, Machlup S. 1953 Fluctuations and irreversible processes. *Phys. Rev.* **91**, 1505-1512. (doi:10.1103/PhysRev.91.1505)
34. Abe S. 2017 Invariants of Fokker-Planck equations. *Eur. Phys. J. Special Topics* **226**, 529-532. (doi:10.1140/epjst/e2016-60215-1)
35. Dirac PAM. 2001 *Lectures on quantum mechanics*. New York, USA: Dover.
36. Holstein BR. 1992 *Topics in advanced quantum mechanics*. Redwood City, USA: Addison-Wesley.
37. Ván P, Kovács R. 2019 Variational principles and thermodynamics. *Philos. T. R. Soc. A* **XXX**, in this volume. (doi:???)
38. Muschik W, Papenfuss C, Ehrentraut H. 2001 A sketch of continuum thermodynamics. *J. Non-Newtonian Fluid Mech*. **96**, 255-290. (doi:10.1016/S0377-0257(00)00141-5)
39. Ván P. 2003 Weakly nonlocal irreversible thermodynamics. *Ann. Phys. (Leipzig)* **12**, 146-173. (doi:10.1002/andp.200310002)
40. Cimmelli VA. 2007 An extension of Liu procedure in weakly nonlocal thermodynamics. *J. Math. Phys*. **48**, 113510. (doi:10.1063/1.2804753)
41. Berezovski A, Ván P. 2017 *Internal variables in thermoelasticity*. Cham, Switzerland: Springer International.